\documentclass[11pt]{article}

\usepackage{amsthm}
\usepackage{algorithm, algorithmic}
\usepackage{graphicx}
\usepackage{amssymb}
\newtheorem{thm}{Theorem}[section]
\theoremstyle{definition}

\theoremstyle{remark}

\theoremstyle{plain}
\newtheorem{lem}[thm]{Lemma}
\newtheorem{col}[thm]{Corollary}

\newtheorem{fig}[figure]{Fig.}

\def\denseformat{
\setlength{\textheight}{9in}
\setlength{\textwidth}{6.9in}
\setlength{\evensidemargin}{-0.2in}
\setlength{\oddsidemargin}{-0.2in}
\setlength{\headsep}{10pt}
\setlength{\topmargin}{-0.3in}
\setlength{\columnsep}{0.375in}
\setlength{\itemsep}{0pt}
}
\def\MathN{\hbox{\rm I\kern-2pt I\kern-3.1pt N}}

\denseformat

\begin{document}

\title{Distributed $(\Delta+1)$-coloring in linear (in $\Delta$) time
}
\author{
Leonid Barenboim\thanks{
        Department of Computer Science,
        Ben-Gurion University of the Negev,
        POB 653, Beer-Sheva 84105, Israel.
        E-mail: {\tt \{leonidba,elkinm\}@cs.bgu.ac.il}
        \newline This research has been supported by the
Israeli Academy of Science, grant 483/06.
        }
         \and
Michael Elkin$^*$ }
\begin{titlepage}
\def\thepage{}
\maketitle
\begin{abstract}
The distributed $(\Delta + 1)$-coloring problem is one of most fundamental and
well-studied problems of Distributed Algorithms. Starting with the work of
Cole and Vishkin in 86, there was a long line of gradually improving
algorithms published. The current state-of-the-art running time is $O(\Delta
\log \Delta + \log^* n)$, due to Kuhn and Wattenhofer, PODC'06. 
Linial (FOCS'87) has proved a lower bound of $\frac{1}{2} \log^* n$ for the problem,
and Szegedy and Vishwanathan (STOC'93) provided a heuristic argument that
shows that algorithms from a wide family of locally iterative algorithms are
unlikely to achieve running time smaller than $\Theta(\Delta \log \Delta)$.

We present a deterministic $(\Delta + 1)$-coloring distributed algorithm with
running time $O(\Delta) + \frac{1}{2} \log^* n$. We also present a tradeoff between
the running time and the number of colors, and devise an $O(\Delta \cdot t)$-coloring algorithm with running time $O(\Delta / t + \log^* n)$, for any parameter $t$, $1 < t \leq \Delta^{1-\epsilon}$, for an arbitrarily small  constant $\epsilon$, $0 < \epsilon < 1$.
Our algorithm breaks the heuristic barrier of Szegedy and Vishwanathan, and
achieves running time which is linear in the maximum degree $\Delta$. On the
other hand, the conjecture of Szegedy and Vishwanathan may still be true, as
our algorithm is not from the family of locally iterative algorithms. 

On the way to this result we study a generalization of the notion of
graph coloring, which is called defective coloring. In an $m$-defective $p$-coloring the
vertices are colored with $p$ colors so that each vertex has up to $m$ neighbors
with the same color. We show that an $m$-defective $p$-coloring with reasonably
small $m$ and $p$  can be computed very efficiently. We also develop a technique
to employ multiple defect colorings of various subgraphs of the original
graph $G$ for computing a $(\Delta+1)$-coloring of $G$. We believe that these
techniques are of independent interest.
\end{abstract}

\end{titlepage}

\section{Introduction}

In the {\em message passing model} of distributed computation \cite{P00} one is given an undirected $n$-vertex graph $G = (V,E)$, whose vertices host processors. The vertices have distinct identity numbers. Each vertex $v$ can communicate with its {\em neighbors}, i.e., vertices $u$ such that $(v,u) \in E$. The communication is synchronous, i.e., it occurs in discrete rounds. Messages are sent in the beginning of each round. A message that is sent in a round $R$, arrives to its destination before the next round $R + 1$ starts. The number of rounds that a distributed algorithm runs is called its {\em running time}.\\ \\
\textbf{1.1 $(\Delta + 1)$-Coloring}\\
Let $\Delta$ denote the maximum degree of $G$. Coloring $G$ with $(\Delta + 1)$ or less colors so that for every pair of neighbors $u$ and $w$, the color of $u$ is different from that of $w$ (henceforth, {\em $(\Delta + 1)$-coloring}) is one of the most central and fundamentally important problems in the area of Distributed Algorithms. In addition to its theoretical appeal, it is very well-motivated by many network primitives that are based on a graph coloring subroutine. (See the introduction of \cite{KW06, SV93} for more details about practical applications.)

The problem has been in the focus of intensive research since mid-eighties. Cole and Vishkin \cite{CV86} devised an $O(\log^* n)$-time 3-coloring algorithm for oriented cycles. In STOC'87 Goldberg and Plotkin \cite{GP87, P88} generalized the algorithm of \cite{CV86} and devised a $(\Delta + 1)$-coloring algorithm that requires $2^{O(\Delta)} + O(\log^* n)$ time. Goldberg, Plotkin and Shannon \cite{GPS88} improved the result of \cite{GP87}, and devised a $(\Delta + 1)$-coloring algorithm with running time $O(\Delta^2 + \log^ * n)$. They have also devised a $(\Delta + 1)$-coloring algorithm with running time $O(\Delta \log n)$. (See also \cite{AGLP89}, FOCS'89, for a more explicit version of the algorithm of \cite{GPS88}.)

In FOCS'87 \cite{L87} Linial devised an $O(\Delta^2)$-coloring algorithm with running time $\log^* n + O(1)$.
Moreover, Linial also proved a lower bound of $\frac{1}{2} \log^* n - O(1)$ for the complexity of the $f(\Delta)$-coloring problem, for any function $f(\cdot)$ \cite{L92}.
In STOC'93 Szegedy and Vishwanathan \cite{SV93} improved the upper bound of \cite{L87}, and devised an $O(\Delta^2)$-coloring algorithm with running time $\frac{1}{2} \log^* n + O(1)$. (See also \cite{MNS95} for a more explicit construction.) Szegedy and Vishwanathan have also presented a heuristic lower bound of $\Omega(\Delta \log \Delta)$ for the complexity of $(\Delta + 1)$-coloring. They considered a class of algorithms that they called "locally iterative algorithms". (See Section 1.3 for more details.) Except for the algorithm of \cite{GPS88} that requires $O(\Delta \log n)$ time, all other $(\Delta + 1)$-coloring algorithms that were known then belong to this family.

Szegedy and Vishwanathan \cite{SV93} presented a heuristic argument that shows that no locally iterative $(\Delta + 1)$-coloring algorithm "is likely to terminate in less than $\Omega(\Delta \log \Delta)$ rounds". More recently, Kuhn and Wattenhofer \cite{KW06} substantiated the heuristic algorithm of \cite{SV93} with a formal proof of a slightly weaker lower bound of $\Omega(\frac{\Delta}{\log^2 \Delta})$ for the class of locally iterative algorithms. Kuhn and Wattenhofer \cite{KW06} have also improved the upper bounds on the complexity of $(\Delta + 1)$-coloring problem, and devised a deterministic algorithm and a randomized algorithm for the problem. The running time of their deterministic (respectively, randomized) algorithm is $O(\Delta \log \Delta + \log^* n)$ (resp., $O(\Delta \log \log n)$).

In this paper we improve upon the state-of-the-art upper bounds of \cite{KW06} on the complexity of $(\Delta + 1)$-coloring problem, and devise a deterministic $(\Delta + 1)$-coloring algorithm with running time $O(\Delta) + \frac{1}{2} \log^* n$. This is the first $(\Delta + 1)$-coloring algorithm with running time linear in $\Delta$. Moreover, our algorithm breaks the heuristic barrier of $\Omega(\Delta \log \Delta)$ due to Szegedy and Vishwanathan \cite{SV93}. On the other hand, the conjecture of Szegedy and Vishwanathan may still be true, as our algorithm is not from the class of locally iterative algorithms. Note also that by the lower bound of Linial \cite{L92}, the second term $\frac{1}{2} \log^* n$ in the running time of our algorithm cannot be improved. See Table 1 for a concise comparison between previous results and our algorithm.

Also, we generalize our result, and devise a tradeoff between the running time of the algorithm and the number of colors it employs. Specifically, for a parameter $t$, $1 < t \leq \Delta^{1-\epsilon}$, for an arbitrarily small constant $\epsilon$, $0 < \epsilon < 1$, a variant of our algorithm computes an $O(\Delta \cdot t)$-coloring within $O(\Delta / t + \log^* n)$ time. \\

\begin{table}
\begin{center}
\begin{tabular}{|l|l|l|l|l|}
\hline

Running time & Reference & & Running Time & Reference \\
\hline
\hline

$2^{O(\Delta)} + O(\log^* n)$        & Goldberg, Plotkin, 
                                        \cite{GP87} & &
$O(\Delta^2) + \frac{1}{2} \log^* n) $         & Szegedy, Vishwanathan, 
                                                  \cite{SV93} \\
\hline
$O(\Delta^2 + \log^* n)$         & Goldberg et al. 
                                   \cite{GPS88} & & 
$O(\Delta \log \Delta + \log^* n) $         &  Kuhn, Wattenhoffer,
                                                \cite{KW06} \\                                
\hline
$O(\Delta \cdot \log n)$         & Goldberg et al. 
                                   \cite{GPS88} & &
$O(\Delta \log \log n)$ rand.        &  Kuhn, Wattenhoffer,
                                                \cite{KW06} \\
\hline
$O(\Delta^2) + \log^* n$         & Linial
                                   \cite{L87}  & &

$O(\Delta) + \frac{1}{2} \log^* n$        & 
{\bf This paper}  \\

\hline
\end{tabular}
\end{center}
\caption[]{
\label{tab1}
A concise comparison of previous $(\Delta + 1)$-coloring algorithms with our algorithm. All listed algorithms except the algorithm of \cite{KW06} that requires $O(\Delta \log \log n)$ time are deterministic.
}
\end{table}
\noindent \textbf{1.2 Maximal Independent Set} \\
A subset $I \subseteq V$ of vertices is called a {\em Maximal Independent Set} (henceforth, MIS) of $G$ if\\
(1) For every pair $u,w \in U$ of neighbors, either $u$ or $w$ do not belong to $I$, and \\
(2) for every vertex $v \in V$, either $v \in I$ or there exists a neighbor $w \in V$ of $v$ that belongs to $I$.

The MIS problem is closely related to the coloring problem, and similarly to the latter problem, the MIS problem is one of the most central and intensively studied problems in Distributed Algorithms. \cite{L86, ABI86, AGLP89, PS95, KMW04}.
Our $(\Delta + 1)$-coloring algorithm gives rise directly to an algorithm with running time $O(\Delta) + \frac{1}{2} \log^* n$ for computing MIS on graphs with maximum degree $\Delta$. Like in the case of the coloring problem, the previous state-of-the-art was the algorithm of Kuhn and Wattenhofer \cite{KW06} that requires $O(\Delta \log \Delta + \log^* n)$ time.

The state-of-the-art randomized algorithms for the MIS problem on general graphs due to Luby \cite{L86} and Alon, Babai and Itai \cite{ABI86}  requires $O(\log n)$ time. The state-of-the-art deterministic algorithm for the problem due to Panconesi and Srinivasan \cite{PS95} requires $2^{c \cdot \sqrt{\log n}}$ time, for some universal constant $c > 0$. Hence for graphs with maximum degree $\Delta = o(\log n)$, our (deterministic) algorithm improves the state-of-the-art (randomized and deterministic) algorithms for the MIS problem. For graphs with $\Delta = o(2^{c \cdot \sqrt {\log n}})$, our algorithm improves the state-of-the-art with respect to deterministic algorithms for the MIS problem. 

Finally, our results give rise directly to improved algorithms for coloring and computing MIS for graphs of bounded arboricity. Specifically, in \cite{BE08} we have shown that graphs of arboricity at most $a$ can be $O(a \cdot t)$-colored in time $O(\frac {a}{t} \log n + a \log a)$, for any parameter $t$, $1 \leq t \leq a$. As argued in \cite{BE08}, this result implies that in $O(a \sqrt{ \log n } + a \log a)$ time one can compute an MIS on these graphs. Our results in the current paper imply an $O(a \cdot t)$-coloring algorithm with running time $O(\frac {a}{t} \log n + a)$, and an algorithm for computing an MIS within $O(a \sqrt{\log n})$ time.\\

\noindent \textbf{1.3 Our Techniques}\\
We study a generalized variant of coloring, called {\em defective coloring}. For a non-negative integer $m$ and a positive integer $\chi$, an {\em $m$-defective $\chi$-coloring} of a graph $G = (V,E)$ is a coloring that employs up to $\chi$ colors and satisfies that for every vertex $v \in V$, there are at most $m$ neighbors of $v$ that are colored by the same color as $v$. Note that the standard notion of $\chi$-coloring corresponds in this terminology to $0$-defective $\chi$-coloring. Defective coloring was introduced by \cite{CCW86}, and extensively studied from graph-theoretic perspective \cite{CGJ97, F93}. However, to the best of our knowledge, we are the first to develop distributed algorithms for computing defective colorings.

We show that $m$-defective $\chi$-colorings for reasonably small values of $m$ and $\chi$ can be efficiently computed in distributed manner. Also, we demonstrate that defective colorings of various appropriate subgraphs of the input graph $G$ can be combined into a $(\Delta + 1)$-coloring of G.
We believe that our technique for computing and employing defective colorings will be useful for improving state-of-the-art bounds for the coloring and the MIS problems on general graphs, and on other important graph families.

Note that our algorithm does not fall into the framework of locally iterative algorithms. In this framework the algorithm starts with computing an initial coloring that may possibly employ many colors, and proceeds iteratively. In each iteration the number of colors is reduced, until no further progress can be achieved. Very roughly speaking, our algorithm partitions the graph to many vertex-disjoint subgraphs, computes defective coloring for each of them, and combines them into a unified $(\Delta + 1)$-coloring of the original graph. The heuristic barrier of $\Omega(\Delta \log \Delta)$ of Szegedy and Vishwanathan \cite{SV93} for locally iterative algorithms suggests that this completely different approach that our algorithm employs is necessary for achieving running time that is linear in $\Delta$ for the $(\Delta + 1)$-coloring problem. \\

\noindent \textbf{1.4 Related Work}\\
Panconesi and Rizzi \cite{PR01} devised yet another $(\Delta + 1)$-coloring algorithm with running time $O(\Delta^2 + \log^* n)$. (In addition to the algorithms of Goldberg et al. \cite{GPS88} and Linial \cite{L92}.)
In SODA'01 De Marco and Pelc \cite{MP01} claimed an $O(\Delta)$-coloring algorithm with running time $O(\log^* n)$. Such a result directly implies a $(\Delta + 1)$-coloring with time $O(\Delta + \log^* n)$. However, unfortunately, their proof contains a gap (see, e.g., \cite{KW06}); once corrected the analysis gives rise to running time of $O(\Delta^2)$, which is known \cite{GPS88, L92, PR01}.
Johansson \cite{J99} devised a randomized $(\Delta + 1)$-coloring algorithm with running time of $O(\log n)$. (If one does not care about message size, the same bound can be achieved by combining the algorithm of Luby \cite{L86} or Alon et al. \cite{ABI86} with Linial's reduction from coloring to MIS \cite{L92}.)

Computing an MIS on graphs with bounded growth was recently intensively studied \cite{KMW05, KMNW05, SW08}. In another recent development, efficient algorithms for coloring and MIS problems for graphs with small arboricity were devised by the authors of the present paper in \cite{BE08}. The main technique in \cite{BE08} is an efficient algorithm for constructing Nash-Williams decomposition distributively, and all other results there rely on this algorithm. However, as shown in \cite{BE08}, constructing Nash-Williams decomposition requires $\Omega(\frac{\log n}{\log \log n})$ time. Consequently, one cannot employ Nash-Williams decomposition to achieve running time of $O(\Delta) + \frac{1}{2} \log^* n$. As discussed above, our algorithms in the present paper rely on completely different ideas. \\

\noindent \textbf{1.5 The Structure of the Paper}\\
In Section 2 we introduce the notation and terminology used throughout the paper. In Section 3 we describe our algorithm for computing defective colorings. In Section 4 we employ the algorithm for defective coloring to devise our $(\Delta + 1)$-coloring algorithm. This algorithm is then used to obtain the tradeoff between the running time and the number of colors. Due to space limitations, all illustrations are delegated to Appendix A. Also, some proofs are delegated to Appendix B.
\section{Preliminaries}
Unless the base value is specified, all logarithms in this paper are of base 2.
For a non-negative integer $i$, the {\em iterative log-function} $\log^{(i)}(\cdot)$ is defined as follows. For an integer $n > 0$, $\log^{(0)} n = n$, and $\log^{(i+1)} n = \log (\log^ {(i)} n)$, for every $i = 0,1,2,...$. Also, $\log^* n$ is defined by: 
$\log^* n = \min \left\{i \  | \  \log^{(i)} n \leq 2 \right\}$. \\
The graph $G'=(V',E')$ is a  {\em subgraph}  of $G=(V,E)$, denoted $G' \subseteq G$, if $V' \subseteq V$ and $E' \subseteq E$. \\
The {\em degree} of a vertex $v$ in a graph $G = (V,E)$, denoted {\em deg(v)}, is the number of edges incident to $v$. A vertex $u$ such that $(u,v) \in E$ is called a {\em neighbor} of $v$ in $G$. For a subset $U \subseteq V$, the degree of $v$ with respect to $U$, denoted {\em deg(v,U)}, is the number of neighbors of $v$ in $U$. The maximum degree of a vertex in $G$, denoted $\Delta(G)$, is defined by $\Delta(G) = \max_{v  \in V} deg(v) $. If the input graph $G$ can be understood from context, we use the notation $\Delta$ as shortcut for $\Delta(G)$. 


A coloring $\varphi: V \rightarrow \MathN$ that satisfies $\varphi(v) \neq \varphi(u)$ for each edge $(u,v) \in E$ is called a {\em legal coloring}.\\
For positive integers $m$ and $p$, a coloring $\varphi': V \rightarrow \left\{ 1,2,...,p \right\}$ that satisfies that for every vertex $v \in V$, the number of neighbors $u$ of $v$ with $\varphi'(u) = \varphi'(v)$ is at most $m$, called an {\em $m$-defective $p$-coloring}. We also say that the graph $G$ is {\em $m$-defective $p$-colored} by $\varphi'$.\\
The {\em defect parameter} of a vertex $v$ with respect to $\varphi'$, denoted $def_{\varphi'}(v)$, is the number of neighbors $u$ of $v$ with $\varphi'(u) = \varphi'(v)$. A defect parameter of a coloring $\varphi$ is defined by $def(\varphi) = \max \left \{ def_{\varphi}(v) \ | \ v \in V \right \}$. 
See Figure 1 for an illustration.\\ 
Some of our algorithms use as a black-box a procedure due to Kuhn and Wattenhofer \cite{KW06}. This procedure accepts as input a graph $G$ with maximum degree $\Delta$, and an initial legal $m$-coloring, and it produces a $(\Delta + 1)$-coloring of $G$ within time $(\Delta + 1) \cdot \left\lceil \log (m /(\Delta + 1)) \right \rceil = O(\Delta \cdot \log (m / \Delta))$. We will refer to this procedure as {\em KW iterative procedure}. The KW iterative procedure is used in \cite{KW06} to devise a $(\Delta + 1)$-coloring algorithm (henceforth, {\em KW algorithm}) with running time $O(\Delta \log \Delta + \log^* n)$.

In all our algorithms we assume that all vertices know the number of vertices $n$, and the maximum degree $\Delta$ of the input graph $G$ before the computation starts. This assumption is required for many coloring algorithms, and in particular, it is required in the algorithms of Linial \cite{L92}, Szegedy and Vishwanathan \cite{SV93}, and Kuhn and Wattenhoffer \cite{KW06}, that are used as black boxes in our algorithm.

Although our distributed model allows sending messages of arbitrary size, all algorithms in this paper employ {\em short} messages, that is, messages with $O(\log n)$ bits each.
\section{Defective Coloring}
\noindent \textbf{3.1 Procedure Refine}\\
In this section we present an algorithm that produces a defective coloring. Many $(\Delta + 1)$-coloring algorithms employ the following standard technique. Whenever a vertex is required to select a color it selects a color that is different from the colors of all its neighbors. Its neighbors select their colors in different rounds. On the other hand, if one is interested in a defective coloring, a vertex can select a color that is used by a few of its neighbors. Moreover, some neighbors can perform the selection in the same round. Consequently, the computation is significantly more efficient than that of $(\Delta + 1)$-coloring, and the number of colors employed is smaller.

We devise a $\left\lfloor \Delta / p\right\rfloor$-defective $p^2$-coloring algorithm. We start with presenting a procedure, called {\em Procedure Refine}, that accepts as input a graph with an $m$-defective $\chi$-coloring, and a parameter $p$, $1 \leq p \leq \Delta$, for some integers $m$, $\chi$, and $p$, and computes $(m + \left\lfloor \Delta/p \right\rfloor)$-defective $p^2$-coloring in time $O(\chi)$.





Suppose that before the invocation of Procedure Refine, the input graph $G$ is colored by an $m$-defective $k$-coloring $\varphi$. For each vertex $v$, let ${\cal S}(v)$ (respectively, ${\cal G}(v)$) denote the set of neighbors $u$ of $v$ that have colors smaller (resp., larger) than the color of $v$, i.e., that satisfy $\varphi(u) < \varphi(v)$ (resp., $\varphi(u) > \varphi(v)$).  Procedure Refine computes a new coloring $\varphi'$. It proceeds in two stages. In the first stage, each vertex $v$ computes a new color $\psi(v)$ from the range $\left\{1,2,...,p\right\}$ in the following way. Once $v$ receives the color $\psi(u)$ from each of its neighbors $u$ from ${\cal S}(v)$, it sets $\psi(v)$ to be the color from $\left\{1,2,...,p\right\}$ that is used by the minimal number of these neighbors, breaking ties arbitrarily. Then, it sends its selection $\psi(v)$ to all its neighbors.
In the second stage, each vertex $v$ computes a new color $\Psi(v)$ from the range $\left\{1,2,...,p\right\}$ in a similar way, except that now it considers only neighbors from ${\cal G}(v)$. Once $v$ receives the color $\Psi(w)$ from each of its neighbors $w$ from ${\cal G}(v)$, it sets $\Psi(v)$ to be the color from $\left\{1,2,...,p\right\}$ that is used by the minimal number of these neighbors. Then, it sends its selection $\Psi(v)$ to all its neighbors.



Once the vertex $v$ has computed both colors $\psi(v)$ and $\Psi(v)$, it sets its final color $\varphi'(v) = (\Psi(v) - 1) \cdot p + \psi(v)$.
Intuitively, the color $\varphi'(v)$ can be seen as a pair $(\Psi(v), \psi(v))$.
This completes the description of Procedure Refine.
Next, we show that the procedure is correct.

\begin {lem}
The coloring $\varphi'$ produced by Procedure Refine is an $(m + \left\lfloor \Delta/p \right\rfloor)$-defective $p^2$-coloring.
\end {lem}
\begin {proof}
First, observe that for each vertex $v$, it holds that $1 \leq \psi(v), \Psi(v) \leq p$, and thus $1 \leq \varphi'(v) \leq p^2$.
It is left to show that for each vertex $v$, the number of neighbors $u$ of $v$ with $\varphi'(u) = \varphi'(v)$ is at most $(m + \left\lfloor \Delta/p\right\rfloor)$. Each vertex $v$ has at most $m$ neighbors $z$ such that $\varphi(v) = \varphi(z)$. By the pigeonhole principle, the number of neighbors $u$ of $v$ with $\varphi(u) < \varphi(v)$ and $\psi(u) = \psi(v)$ is at most $\left\lfloor \left| {\cal S}(v) \right| / p\right\rfloor$, since $v$ selected $\psi(v)$ to be the color from $\left\{ 1,2,...,p \right\}$
that is employed by the minimal number of neighbors from ${\cal S}(v)$. Similarly, the number of neighbors $w$ of $v$ with $\varphi(w) > \varphi(v)$ and $\Psi(w) = \Psi(v)$ is at most $\left\lfloor \left| {\cal G}(v) \right| / p\right\rfloor$. Observe that for any neighbor $u$ of $v$, if $\varphi'(u) = \varphi'(v)$ then $\psi(u) = \psi(v)$ and $\Psi(u) = \Psi(v)$. Consequently, the number of neighbors $u$ with $\varphi'(u) = \varphi'(v)$ is at most $(m + \left\lfloor \left| {\cal S}(v) \right| / p\right\rfloor + \left\lfloor \left| {\cal G}(v) \right| / p\right\rfloor) \leq (m + \left\lfloor deg(v) / p \right\rfloor) \leq (m + \left\lfloor \Delta / p \right\rfloor)$.
\end {proof}


\def\APPb{
\begin {lem} \label{reftime}
The time complexity of Procedure Refine is $\chi + 1$.
\end {lem}
\textit{Proof.}
We prove by induction on $i$ that after $i$ rounds, $i = 1,2,...,\chi$, each vertex with $\varphi(v) \leq i$ has selected its color $\psi(v)$. For the base case, consider all the vertices $v$ with $\varphi(v) = 1$. There are no vertices $u$ with $\varphi(u) < 1$, and thus, each vertex $v$ with $\varphi(v) = 1$ selects the color $\psi(v)$ in the first round. Now, assume that after $(i - 1)$ rounds, each vertex with $\varphi(v) \leq (i - 1)$ has selected its color $\psi(v)$. Then, by the induction hypothesis, in round $i$, for a vertex $v$ with $\varphi(v) = i$, all the neighbors $u$ of $v$ that satisfy $\varphi(u) < \varphi(v) = i$ have selected their color $\psi(u)$ in round $(i-1)$ or earlier. Hence, if $v$ has not selected the color $\psi(v)$ before round $i$, it necessarily selects it on round $i$. Therefore, after $\chi$ rounds all the vertices in the graph have selected the color $\psi(v)$ and the first stage is completed. Similarly, the second stage is completed after another $\chi$ rounds. The computation of $\varphi'(v)$ from $\psi(v)$ and $\Psi(v)$ is performed immediately after the second stage is finished, and it requires no additional communication. The total running time of the procedure is, therefore, $O(\chi)$.
Finally, note that the two stages can be executed in parallel. Thus, the required running time is $\chi + 1$.
}

The two stages of Procedure Refine can be executed in parallel. Consequently, Procedure Refine can be executed within $\chi + 1$ rounds. (See Lemma \ref{reftime} in Appendix B for a formal proof.) We summarize this section with the following corollary.

\begin{col} \label{sumref}
For positive integers $\chi$, $m$, and $p$, suppose that Procedure Refine is invoked on a graph $G$ with maximum degree $\Delta$. Suppose also that $G$ is $m$-defective $\chi$-colored. Then the procedure produces an $(m + \left\lfloor \Delta /p \right \rfloor)$-defective $p^2$-coloring of $G$. It requires at most $\chi + 1$ rounds.
\end{col}

\noindent \textbf{3.2 Procedure Defective-Color}\\
In this section we devise an algorithm called {\em Procedure Defective-Color}. The algorithm accepts as input a graph $G = (V,E)$, and two integer parameters  $p, q$ such that $1 \leq p \leq \Delta$, $p^2 < q$, and $q < c' \cdot \Delta^2$, for some positive constant $c' > 0$. It computes an $O(\frac{\log \Delta} {\log (q / p^2)} \cdot \Delta / p)$-defective $p^2$-coloring of $G$ in time $O(\log^* n + \frac{\log \Delta} {\log (q / p^2)} \cdot q)$ from scratch.
In particular, if we set $q = \Delta^{\epsilon} \cdot p^2$ for an arbitrarily small positive constant $\epsilon$, we get an $O(\Delta / p)$-defective $p^2$-coloring algorithm with running time $O(\log^* n + \Delta^{\epsilon} \cdot p^2)$.

The algorithm starts by computing an $O(\Delta^2)$-coloring of the input graph. This coloring $\varphi$ can be computed in $O(\log^* n)$ time from scratch using the algorithm of Linial \cite{L92}. In some scenarios in which the procedure accepts some auxiliary coloring of $G$ as part of the input, one can compute an $O(\Delta^2)$-coloring much faster. The latter case is described in detail in Section 4. Let $c$, $c > 0$, be a constant such that $c \cdot (\Delta^2)$ is an upper bound on the number of colors employed. 
Let $h = \left\lfloor c \cdot \Delta^2 / q \right\rfloor$. (The constant $c'$ is sufficiently small to ensure that $h \geq 1$). Each vertex $v$ with $1 \leq \varphi(v) \leq h \cdot q$ joins the set $V_j$ with $j = \left\lceil \varphi(v) / q \right \rceil$. Vertices $v$ that satisfy $h \cdot q < \varphi(v) \leq c \cdot \Delta^2$ join the set $V_h$. In other words, the index $j$ of the set $V_j$ to which the vertex $v$ joins is determined by 
 $j = \min \left\{ \left\lceil \varphi(v) / q \right\rceil, h \right\}$. Observe that for every index $j$, $1 \leq j \leq h - 1$, the set $V_j$ is colored with exactly $q$ colors, and $V_h$ is colored with $q'$ colors with $ q \leq q' \leq 2q$. By definition, for each $j$, $1 \leq j \leq h - 1$, $V_j$ is $0$-defective $q$-colored (i.e., $m= 0$, $k = q$), and $V_h$ is $0$-defective $q'$-colored ($m = 0$, $k = q'$). For each $j$, $1 \leq j \leq h$, we denote this coloring of $V_j$ by $\psi_j$. Then, for each graph $G(V_j)$ induced by the vertex set $V_j$, Procedure {\em Refine} is invoked on $G(V_j)$ with the parameter $p$, in parallel for $j = 1,2,..,h $. As a result of these invocations, each graph $G(V_j)$ is now $\left\lfloor \Delta / p \right \rfloor$-defective $p^2$-colored. Let $\varphi'_j$ denote this coloring. Next, each vertex $v$ selects a new color $\varphi''(v)$ by setting $\varphi''(v) = \varphi'_j(v) + (j - 1) \cdot p^2$. The number of colors used by the new coloring $\varphi''$ is at most $h \cdot p^2 \leq c \cdot (\Delta^2) \cdot p^2 / q$. Later we argue that the coloring $\varphi''$ is a $\left\lfloor  \Delta/p \right\rfloor$-defective $(c \cdot (\Delta^2) \cdot p^2 / q)$-coloring of the input graph $G$.

This process is repeated iteratively. On each iteration the vertex set is partitioned into disjoint subsets $V_j$, such that in each subset the vertices are colored by at most $q$ different colors, except one subset in which the vertices are colored by at most $2q$ colors. Then, in parallel, the coloring of each subset is converted into $p^2$-coloring. Consequently, in each iteration the number of colors is reduced by a factor of at least $q / p^2$. (Except the last iteration in which the number of colors is larger than $p^2$ but smaller than $2q$, and it is reduced to $p^2$.) However, for a vertex $v$, the number of neighbors of $v$ that are colored by the same color as $v$, $def_{\varphi}(v)$, may grow by an additive term of $\left\lfloor \Delta/p\right\rfloor$ in each iteration. The process terminates when the entire graph $G$ is colored by at most $p^2$ colors. (After $\log_{q/p^2} c \cdot \Delta^2$ iterations all vertices know that $G$ is colored by at most $p^2$ colors.)  In each iteration an upper bound $\chi$ on the number of currently employed colors is computed. In the last iteration, if $\chi < q$ then all the vertices join the same set $V_1$, and consequently $V_1 =  V$, and Procedure Refine is invoked on the entire graph $G$. The pseudo-code of the algorithm is provided below. See Figure 2 in Appendix A for an illustration.

\begin{algorithm}[H]
\caption{Procedure Defective-Color($p, q$) (Algorithm for a vertex $v$)
}
\label{proced:relaxed}

\begin{algorithmic}[1] 

\STATE $\varphi$ := color $G$ with $c \cdot (\Delta^2)$ colors

\STATE $\chi$ := $c \cdot (\Delta^2)$ \ \ \ \ \ /* the current number of colors */

\STATE $i = 0$ \ \ \ \ \ \ \ \ \ \ \ \ \ \ \ /* the index the of current iteration */

\WHILE {$\chi > p^2$}

  \IF {$\chi < q$}
         
         \STATE $j : = 1$
         
  \ELSE

   \STATE $j$ := $\min \left\{ \left\lceil \varphi(v) / q \right\rceil, \left\lfloor \chi / q \right\rfloor \right\}$
      
   \ENDIF
   
   \STATE set $V_j$ to be the set of $v$
   
   \STATE $\psi_j(v) := \varphi(v) - (j - 1) \cdot q$ \ \ \ \ \ \ \ /* $\psi_j(v)$ is an $(i \cdot \left\lfloor \Delta / p \right\rfloor)$-defective $(2q)$-coloring of $G(V_j)$ */
   
   \STATE $\varphi'_j$ := invoke Procedure Refine on $G(V_j)$ with the coloring $\psi_j$ and the parameter $p$ as input
   
   \STATE $\varphi(v) := \varphi''(v)$ := $\varphi'_j(v) + (j - 1) \cdot p^2$
   
   \STATE $\chi$ := $(\max \left\{ \left\lfloor  \chi / q \right\rfloor, 1 \right\}) \cdot p^2$   \ \ \ \ \ \ \ \ \ \ /* $\varphi(v)$ is an $(i \cdot \left\lfloor \Delta / p \right \rfloor)$-defective $\chi$-coloring of $G$ */
   
   \STATE $i := i + 1$
   
\ENDWHILE

\STATE return $\varphi$

\end{algorithmic}
\end{algorithm}







   
   
   
   
   
   



In what follows we prove the correctness of Procedure Defective-Color. We start with proving the following invariant regarding the variable $\chi$. Let $\chi_i$ denote the value of $\chi$ at the end of the $i$th iteration. For technical convenience, we define $\chi_0$ to be the value of $\chi$ at the beginning of the first iteration.

\begin{lem} \label{colorsk}
For $i= 0,1,2,...$, after the $i$th iteration, the number of colors employed by $\varphi$ is at most $\chi_i$.
\end{lem}
\def\APPc{
The proof is by induction on $i$.\\
\textbf{Base $(i = 0)$:} In the first step of Procedure Defective-Color, the graph $G$ is colored using  $c \cdot (\Delta^2)$ colors. Therefore, after 0 iterations, the number of colors employed by $\varphi$ is at most $k_0 = c \cdot (\Delta^2)$.\\
\textbf{Induction step:} By the induction hypothesis, after iteration $(i - 1)$, the number of colors employed by $\varphi$ is at most $\chi_{i-1}$. In iteration $i$ the vertex set $V$ of $G$ is partitioned into $\max \left\{ \left\lfloor \chi_{i-1} / q \right\rfloor , 1 \right\}$ disjoint subsets $V_j$, $j = 1,2,...,\max \left\{ \left\lfloor \chi_{i-1} / q \right\rfloor, 1 \right\}$. Let $h = \left\lfloor \chi_{i-1} / q \right \rfloor$. Each of these subsets except $V_h$ is colored with $q$ colors. If $h > 0$, then $V_h$ is colored with at least $q$ colors, but no more than $2q$ colors. Procedure Refine produces a new coloring in each set $V_j$ such that the number of colors used in the set $V_j$ is $p^2$, for $j = 1,2,..., \max \left\{ \left\lfloor \chi_{i-1} / q \right\rfloor, 1 \right\}$. Consequently, the number of colors used by $\varphi$ at the end of iteration $i$ is at most $(\max \left\lfloor \left\{ \chi_{i-1} / q \right\rfloor , 1 \right \}) \cdot p^2 = \chi_i$. (See step 14 of Algorithm 1.)
}
The proof of Lemma \ref{colorsk} appears in Appendix B.\\
By step 14 of Algorithm 1, $\chi_{i+1} \leq \max \left\{ \chi_i \cdot p^2 / q, \ p^2 \right \}$, for $i = 0,1,2,...$, and $\chi_0 = c \cdot \Delta^2$. Therefore, \\ $\chi_i \leq \max \left\{c \cdot \Delta^2 \cdot (p^2 /q )^i, \ p^2 \right\}$. For future reference, this fact is summarized in the following corollary.

\begin{col} \label{relit}
For $i = 0,1,2,...$, the number of colors employed by $\varphi$ in the beginning of iteration $(i + 1)$ is $\chi_i \leq \max \left\{c \cdot \Delta^2 \cdot (p^2 /q )^i, \ p^2 \right\}$.
\end{col}

Next, we analyze the defect parameter of the coloring produced by Procedure Defective-Color.

\begin{thm} \label{relcor}
Procedure Defective-Color invoked with the parameters $p, q$, computes an $O(\frac{\log \Delta} {\log (q / p^2)} \cdot \Delta / p)$-defective $p^2$-coloring.
\end{thm}  
\begin{proof}
We prove by induction on $i$ that after $i$ iterations $\varphi(v)$ is an  $(i \cdot \Delta / p)$-defective \\ $(\max \left \{ c \cdot \Delta^2 \cdot (p^2 / q)^i, \ p^2 \right \})$-coloring of $G$.\\
\textbf{Base $(i = 0)$:} Observe that a $0$-defective $(c \cdot \Delta^2)$-coloring is computed in the first step of the algorithm. Therefore, before the begining of the first iteration, $\varphi$ is a $0$-defective $(c \cdot \Delta^2)$-coloring of $G$. \\
\textbf{Induction step:} Let $\varphi$ be the coloring produced after $i - 1$ iterations. By the induction hypothesis, $\varphi$ is an $((i - 1) \cdot \Delta / p)$-defective ($\max \left\{c \cdot \Delta^2 \cdot (p^2 / q)^{i - 1}, \ p^2\right\})$-coloring of $G$. In iteration $i$, the vertex set $V$ of $G$ is partitioned into $\max \left\{ \left\lfloor \chi_{i-1} / q \right\rfloor, 1 \right\}$ disjoint subsets $V_j$. If there is only one subset $V_1 = V$, then $G(V_1) = G$ is colored with at most $2q$ colors. Otherwise, each induced graph $G(V_j)$, $1 \leq j <  \left\lfloor \chi_{i-1} / q \right\rfloor$, is colored by $q$ different colors. The induced graph $G(V_{\left\lfloor \chi_{i-1} / q \right\rfloor})$ is colored by at most $2q$ colors. Therefore, for each $j$, $1 \leq j \leq \max \left \{ \left\lfloor \chi_{i-1} / q \right \rfloor, 1 \right\}$ the coloring $\psi_j$ computed in step 11 of the $i$th iteration is an $((i - 1) \cdot \Delta / p)$-defective $(2q)$-coloring of $G(V_j)$. In step 12, Procedure Refine is invoked on $G(V_j)$ with $p$ as input. As a result, an $((i - 1) \cdot \Delta / p + \Delta/p)$-defective $p^2$-coloring $\varphi'_j$ of $G(V_j)$ is produced. In other words $\varphi'_j$ is an $(i \cdot \Delta / p)$-defective $p^2$-coloring of $G(V_j)$, i.e., $def(\varphi'_j) \leq i \cdot \Delta / p$. 

Consider a vertex $v$, and a neighbor $u$ of $v$. First, suppose that $v \in V_j$, $u \in V_{\ell}$, and $j < \ell$. Then $\varphi''(v) - \varphi''(u)$ $=(\varphi'_j(v) - \varphi'_{\ell}(u)) + (j - \ell) \cdot p^2 \geq \varphi'_j(v) - \varphi'_{\ell}(u) + p^2$. Since $\varphi'_j(v) - \varphi'_{\ell}(u) \geq - p^2 + 1$, it follows that $\varphi''(v) \neq \varphi''(u)$.
Second, consider a neighbor $w \in V_j$ of $v$.
If $\varphi'_j(v) \neq \varphi'_j(w)$ then also $\varphi''(v) = \varphi'_j(v) + (j - 1) \cdot p^2 \neq \varphi''(w) = \varphi'_j(w) + (j - 1) \cdot p^2$. Since $def(\varphi'_j) \leq i \cdot \Delta / p$, there are at most $(i \cdot \Delta / p)$ neighbors $w \in V_j$ of $v$ such that $\varphi'_j(w) = \varphi'_j(v)$. Consequently, the coloring $\varphi = \varphi''$ that is produced in step 13 of the $i$th iteration is an $(i \cdot \Delta / p)$-defective $(\max \left \{ c \cdot \Delta^2 \cdot (p^2 / q)^i, \ p^2 \right \})$-coloring of $G$. This completes the inductive proof.
By Corollary \ref{relit}, after $\frac{\log (c \cdot \Delta^2)} {\log (q / p^2)}$ iterations, $\varphi$ is a $(\frac{\log (c \cdot \Delta^2)} {\log (q / p^2)} \cdot \Delta / p)$-defective $p^2$-coloring of $G$. 
\end{proof}

Procedure Defective-Color starts with computing an $O(\Delta^2)$-coloring. The algorithm of Linial \cite{L92} computes a $(c \cdot \Delta^2)$-coloring in time $\log^* n + O(1)$.
Szegedy and Vishwanathan \cite{SV93} showed that the coefficient of $\log^* n$ can be improved to 1/2, i.e., they devised an $O(\Delta^2)$-coloring algorithm with time $\frac{1}{2} \log^* n + O(1)$. Henceforth we refer to this algorithm as the {\em SV algorithm}.
The number of iterations performed by Procedure Defective-Color is at most $\log_{q/ p^2} (c \cdot \Delta^2) = \frac{\log (c \cdot \Delta^2)} {\log (q / p^2)}$. Each iteration invokes Procedure Refine that requires $O(q)$ time, and performs some additional computation that requires $O(1)$ time. The running time of Procedure Defective-Color is given in the following theorem.

\begin{thm} \label{reltime}
Procedure Defective-Color invoked with parameters p,q, runs in $T(n) + O(q \cdot \frac{\log \Delta}{\log (q / p^2)})$ time, where $T(n)$ is the time required for computing $O(\Delta^2)$-coloring. If the SV algorithm is used for $O(\Delta^2)$-coloring, the running time of Procedure Defective-Color is
$O(q \cdot \frac{\log \Delta}{\log (q / p^2)}) + \frac{1}{2} \log^* n$.
\end{thm}

\section{$(\Delta + 1)$-Coloring}
In this section we employ the techniques and algorithms described in Section 3 to devise an efficient $(\Delta + 1)$-coloring algorithm.
As a first step, we devise a $(\Delta + 1)$-coloring algorithm ${\cal J}$ with running time $O(\Delta \log \log \Delta) + \log^* n$. Set $p = \log \Delta$, and $q = \Delta^{\epsilon}$, for an arbitrarily small positive constant $\epsilon$, $0 < \epsilon < 1$. By Theorems \ref{relcor} and \ref{reltime}, Procedure Defective-Color invoked with these parameters computes an $O(\Delta / \log \Delta)$-defective $(\log^2 \Delta)$-coloring $\varphi$ in $O(\Delta^{\epsilon}) + \frac{1}{2} \log^* n$ time. 
Let $V_j$ denote the set of vertices $v$ with $\varphi(v) = j$, for $j = 1,2,...,\left\lfloor \log^2 \Delta \right \rfloor$. Observe that the maximum degree $\Delta_j = \Delta(G(V_j))$ of the graph $G(V_j)$ induced by $V_j$ is at most the defect parameter $def(\varphi)$ of the coloring $\varphi$. Thus, $\Delta_j = O(\Delta / \log \Delta)$. Consequently, all graphs $G(V_j)$ can be colored in parallel with $O(\Delta / \log \Delta)$ colors each using the KW algorithm. The running time of this step is $O(\Delta + \log^* n)$. If we use distinct palettes of size $O(\Delta / \log \Delta)$ for each graph $G(V_j)$, then we get an $O(\log^2 \Delta \cdot \Delta / \log \Delta) = O(\Delta \log \Delta)$-coloring of the entire graph $G$. Next, we use the KW iterative procedure with the parameter $m = O(\Delta \log \Delta)$ to compute a $(\Delta + 1)$-coloring from $O(\Delta \log \Delta)$-coloring in time $O( \Delta \cdot \log \frac{m}{\Delta}) = O(\Delta \log \log \Delta)$. The total running time of the above algorithm for computing $(\Delta + 1)$-coloring is $O(\Delta \log \log \Delta +  \log^* n)$.

\begin{col} \label{deltaf}
The algorithm ${\cal J}$ computes a $(\Delta + 1)$-coloring in time $O(\Delta \cdot \log \log \Delta +  \log^* n)$.
\end{col}

Corollary \ref{deltaf} is already a significant improvement over the previous state-of-the-art running time of $O(\Delta \cdot \log \Delta + \log^* n)$, due to Kuhn and Wattenhofer \cite{KW06}. In what follows we improve this bound further, and devise a $(\Delta + 1)$-coloring algorithm with running time $O(\Delta) + \frac{1}{2} \log^* n$. We do it in two steps. First, we improve it to $O(\Delta \cdot \log^{(k)} \Delta + \log^* n)$, for an arbitrarily large constant integer $k$. Second, we achieve our ultimate goal of $O(\Delta) + \frac{1}{2} \log^* n$.

Suppose that there exists an algorithm ${\cal A}_k$ that computes a $(\Delta + 1)$-coloring in $O(\Delta \log^{(k)} \Delta) + \frac{k}{2} \cdot \log^* n$ time, for some integer $k > 0$. We employ this algorithm to devise a more efficient $(\Delta + 1)$-coloring algorithm ${\cal A}_{k+1}$. Specifically, ${\cal A}_{k+1}$ has running time $O(\Delta \log^{(k + 1)} \Delta) + \frac{(k + 1)}{2} \cdot \log^* n$. For an input graph $G$, invoke Procedure Defective-Color with the parameters $p = \log^{(k)} \Delta$, $q = \Delta^{\epsilon}$, for a constant $\epsilon$, $0 < \epsilon < 1$. We obtain an $O(\Delta / \log^{(k)} \Delta)$-defective $((\log^{(k)} \Delta)^2)$-coloring of $G$, and the running time of this step is $O(\Delta^{\epsilon}) + \frac{1}{2} \log^* n$. Let $V_j$ denote the subset of vertices that were assigned the color $j$. Invoke in parallel the algorithm ${\cal A}_k$ on the subgraphs $G(V_j)$, for $j = 1,2,...,p^2$, using distinct palettes. The resulting coloring of these invocations is a (0-defective) $O(\Delta \log^{(k)} \Delta)$-coloring. Invoke the KW iterative procedure with the parameter $m = O(\Delta \log^{(k)} \Delta)$ to compute a $(\Delta + 1)$-coloring of $G$ in time $O(\Delta \log \frac{m}{\Delta}) = O(\Delta \log^{(k+1)} \Delta)$. 

The running time of the algorithm ${\cal A}_{k+1}$ consists of the running time of Procedure Defective-Color, which is $O(\Delta^{\epsilon}) + \frac{1}{2} \log^* n$, the running time of the algorithm ${\cal A}_k$ on graphs with maximal degree $\Delta / \log^{(k)} \Delta$, which is $O(\Delta) + \frac{k}{2} \cdot \log^* n$, and the running time of the KW iterative procedure which is $O(\Delta \log^{(k+1)}\Delta)$. Therefore, the total running time of ${\cal A}_{k+1}$ is $O(\Delta \log^{(k+1)}\Delta) + \frac{(k+1)}{2} \cdot \log^* n$. 


We summarize this argument  with the following theorem.
\begin{thm} \label{iterk}
For a constant arbitrarily large positive integer $k$, the algorithm ${\cal A}_k$ computes a $(\Delta + 1)$-coloring of the input graph in time $O(\Delta \log^{(k)} \Delta + \log^* n)$.
\end{thm}

Next, we demonstrate that by a slight change of the algorithm and more careful analysis one can improve the running time even further, and achieve running time of $O(\Delta) + \frac{1}{2} \log^* n$.

The algorithm ${\cal A}_k$ starts with invoking Procedure Defective-Color, which partitions the vertex set of $G$ into disjoint subsets $V_1,V_2,...$. Then it invokes the algorithm ${\cal A}_{k-1}$ on each of the subsets. Essentially, this step is a recursive invocation of our algorithm, and the depth of the recursion is $k$. Finally, it invokes the KW iterative procedure to merge the colorings that the recursive invocations return into a unified coloring of the entire graph $G$.

Procedure Defective-Color is invoked on each of the $k$ levels of recursion. (Moreover, in all except the highest level it is invoked many times, but these invocations occur in parallel.)
Each of these invocations entails an invocation of the SV algorithm, which requires $\frac{1}{2} \log^* n$ time for each invocation. Next, we argue that one can save time and use just one single invocation of the SV algorithm.

In the modified variant of our algorithm we invoke the SV algorithm just once, in the very beginning of the computation. Let $\lambda$ denote the resulting $(c \cdot \Delta^2)$-coloring, for some explicit positive integer $c$. 
Then, each time Procedure Defective-Color has to compute a $(c \cdot \Delta(G')^2)$-coloring for a subgraph $G' \subseteq G$, instead of invoking the SV algorithm it employs the following technique. This technique computes the desired coloring in a one single round, based on the coloring $\lambda$. It is based on the following theorem.

\begin{thm} \label{sqcol} \cite{EFF85, L92} :
For every positive integer A, there exists a collection ${\cal T}$ of  $\Theta(A^3)$ subsets of $\left\{ 1,2,...,c \cdot A^2 \right \}$ such that for every $A + 1$ subsets $T_0,T_1,...,T_A \in {\cal T}$, $T_0 \nsubseteq \bigcup_{i=1}^A T_i$.
\end{thm}

Consider a subgraph $G'$ of $G$, and let $\Delta'$ be an upper bound on the maximum degree of $G'$ that satisfies that $\Delta' = \Omega(\frac{\Delta}{\log \Delta})$. Then given a $(c \cdot \Delta^2)$-coloring $\lambda$ of $G$, we compute a $(c \cdot (\Delta')^2)$-coloring $\lambda'$ of $G'$ in the following way.
Set $A = \Delta'$, and 
let $\cal T$ be the collection whose existence is guaranteed by Theorem \ref{sqcol}. We assign a distinct subset from ${\cal T}$ to each color of $\lambda$. 
(Note that the number of subsets, that is, $\Theta(A^3)$ is greater than the number of colors $c \cdot \Delta^2 = O(A^2 \log^2 A)$ that $\lambda$ employs.)
For a vertex $v \in V(G')$, let $T \in {\cal T}$ denote the subset assigned to $\lambda(v)$. Let $\widetilde{T}$ denote the union of subsets assigned to the colors of the neighbors of $v$. Since $v$ has at most $A = \Delta'$ neighbors, by Theorem \ref{sqcol}, there exists a member $t \in T$ such that $t \notin \widetilde{T}$. The vertex $v$ selects $t$ as its new color. Since all the neighbors of $v$ select their new colors from $\widetilde{T}$, the resulting coloring is a legal $O(A^2)$-coloring.

This process requires one round. The collection $\cal{T}$ is computed locally by each vertex with no communication whatsoever. In a single round, each vertex $v \in G'$ sends its color $\lambda(v)$ to all its neighbors in $G'$. Once $v$ knows the colors $\lambda(u)$ of all its neighbors $u \in G'$, it computes locally the sets $T$ and $\widetilde{T}$, and selects a new color from the set $T \setminus \widetilde {T}$. (This set is necessarily not empty, by Theorem \ref{sqcol}.) 
Since $T \subseteq \left\{1,2,...,O((\Delta')^2) \right\}$, the process computes a legal $O((\Delta')^2)$-coloring in a single round.

The pseudo-code of the procedure for computing $(\Delta + 1)$-coloring in time $O(\Delta) + \frac{1}{2} \log^* n$, Procedure {\em Delta-Color}, is given below. 
The procedure accepts as input a coloring $\psi$, a positive integer parameter $i$ that reflects the recursion level, and a parameter $\Lambda$. The parameter $\Lambda$ is an upper bound on the maximum degree of the graph $G$. In the very first invocation, the parameter $\psi$ is set as $\lambda$, and the parameter $\Lambda$ is set as $\Delta$. In step 6 Procedure defective-Color is invoked. However, actually the invoked procedure is slightly different from Procedure Defective-Color, 
that is described in Algorithm 1. Specifically, while in Algorithm 1 on step 1 the algorithm of Linial (or the SV algorithm) is invoked to compute the coloring $\varphi$, here we invoke the SV algorithm to compute the coloring $\lambda$ before invoking Procedure Delta-Color for the very first time. In its first invocation Procedure Defective-Color uses the coloring $\lambda$ that the first invocation of Procedure Delta-Color received as a part of its input, and sets $\varphi := \lambda$ on its step 1. In all consequent invocations  of Procedure Defective-Color the respective colorings $\varphi_j$ computed in step 9 of Algorithm 2 are used, i.e., the procedure sets $\varphi := \varphi_j$ on its step 1. These colorings are computed using Theorem \ref{sqcol}, as described above.

\begin{algorithm}[H]
\caption{Procedure Delta-Color($G, \Lambda, i, \psi$)
}
\label{proced:delta}
\begin{algorithmic}[1]
\IF {i = 1}
     
     \STATE compute a $(\Lambda + 1)$-coloring of $G$ from $\psi$ using the KW iterative procedure

\ELSE

     \STATE $k$ := $\left\lfloor \log^{(i-1)} \Lambda\right\rfloor$
     
     \STATE $d$ := $\left\lfloor \Lambda / k \right\rfloor$

     \STATE $\varphi$ :=  Procedure Defective-Color($p := k$, $q := \left\lfloor \Lambda^{\epsilon}\right \rfloor$)  
     
     \STATE let $V_j$, $j = 1,2,...,k^2$, denote the set of vertices such that $\varphi(v) = j$
          
     \FOR { $j = 1,2,...,k^2$, in parallel} 
     
          \STATE  $\varphi_j$ := compute $O(d^2)$-coloring of $G(V_j)$ using Theorem \ref{sqcol}
     
          \STATE $\varphi'_j$ := invoke Procedure Delta-Color($G(V_j)$, $d$, $i-1$, $\varphi_j$) recursively 
          
          \FOR {each $v \in G(V_j)$, in parallel} \STATE $\psi(v) := \varphi'_j(v) + (d + 1)(j - 1)$
          
          \ENDFOR
          
     \ENDFOR
     
     \STATE compute a $(\Lambda + 1)$-coloring of $G$ from $\psi$ using the KW iterative procedure   
  
\ENDIF

\end{algorithmic}
\end{algorithm}

     


     
          
     
     
  



Observe that each vertex $v \in V$ has to know an upper bound on the maximal degree of the subgraph it belongs to. Initially, $v$ belongs to $G$ and knows that the maximal degree of $G$ is $\Lambda = \Delta(G)$. Before Procedure Delta-Color is invoked recursively with depth parameter $i-1$, the vertex $v$ computes (step 5 of Algorithm 2) an upper bound on the maximal degree of $G(V_j)$, given by $d = \left\lfloor \Delta(G) / \left\lfloor \log^{(i-1)} \Delta(G)\right\rfloor \right\rfloor$, and passes it as a parameter to the recursive invocation of Procedure Delta-Color on the subgraph $G(V_j)$. 

In the following theorem, we prove the correctness of the algorithm.

\begin{thm} \label{dcor}
If Procedure Delta-Color is invoked on an input graph $G$ with an $O(\Delta^2)$-coloring $\lambda$, an integer parameter $i > 0$, and the maximum degree $\Delta$, it computes a $(\Delta + 1)$-coloring of $G$.
\end{thm}
\begin{proof}
The proof is by induction on $i$.\\
\textbf{Base $(i = 1)$:} In this case the KW iterative procedure is executed on $G$. The correctness follows from the correctness of the KW iterative procedure. \\
\textbf{Induction step:} Suppose that the procedure is invoked with a parameter $i > 1$. In step 6 an $O(d)$-defective $k^2$-coloring $\varphi$ is computed. In step 9, for $j = 1,2,...,k^2$, an $O(d^2)$-coloring of $G(V_j)$ is computed, where $d$ is an upper bound on the maximal degree of $G(V_j)$. By the induction hypothesis, the coloring $\varphi'$ that is computed in step 10 is a $(d + 1)$-coloring of $G(V_j)$.  For any pair of neighbors $u,v$ in $G$, $v \in V_j$, $u \in V_\ell$, it holds that $\psi(v) - \psi(u) =  \varphi'_j(v) + (d + 1)(j - 1) - (\varphi'_{\ell}(u) + (d + 1)(\ell - 1))$. If $j = \ell$ then $\varphi'_j(v) \neq \varphi'_{\ell}(u)$. Otherwise, suppose without loss of generality that $j > \ell$. Thus, $\psi(v) - \psi(u) \geq (d + 1) + (\varphi'_j(v) - \varphi'_{\ell}(u)) \geq 1$. Hence $\psi(v) \neq \psi(u)$, and thus the coloring $\psi$ that is computed in step 12 is a legal $(k^2 \cdot (d + 1))$-coloring. 
Recall that $k = \left\lfloor \log^{(i-1)} \Lambda \right \rfloor$, $d = \left\lfloor \Lambda / k \right \rfloor$, and the parameter $\Lambda$ is set as $\Delta$.
Therefore, $\psi$ is a legal $O(\Delta \log^{(i-1)} \Delta)$-coloring, and consequently the coloring computed in step 15 by KW coloring algorithm is a legal $(\Delta + 1)$-coloring. 
\end{proof}

Next, we analyze the running time of Procedure Delta-Color.
Let $c$, $c \geq 2$, denote a universal constant such that $T(n) + c \cdot \Delta^{\epsilon}$ is an upper bound on the running time of Procedure Defective-Color (see Theorem \ref{reltime}), and $c \cdot \Delta^2$ is an upper bound on the number of colors employed by the SV algorithm. (It is easy to verify that in Theorem \ref{reltime}, $c = O(\epsilon^{-1})$. However, $\epsilon > 0$ is a universal constant.)

\begin{lem} \label {druntime}
The running time of Procedure Delta-Color invoked on the input graph $G$ with a $(c \cdot \Delta^2)$-coloring $\lambda$, and an integer parameter $i$, $0 < i \leq \log^* \Delta$, is at most
 $$\tau(i,\Delta) = i + i \cdot c \cdot \Delta^{\epsilon} + c \cdot \sum_{j = 0}^i \frac{1}{2^j} \Delta + c \cdot \sum_{j = 0}^{i-1} \log^{(j)}\Delta+ c \cdot (\Delta + 1) \log^{(i)} \Delta = (c  + 2 + o(1)) \cdot (\Delta + 1) \cdot \log^{(i)} \Delta.$$
\end{lem}
The proof can be found in Appendix B. \\
\def\APPd{

\noindent The proof is by induction on $i$.\\
\textbf{Base $(i = 1)$:} In this case, the KW iterative procedure is invoked on the graph $G$ with the $(c \cdot \Delta^2)$-coloring $\lambda$. The running time is $(\Delta + 1) \left\lceil \log (c \cdot \Delta) \right \rceil \leq 1 + \Delta + (\Delta + 1) \log (c \cdot \Delta)$, and it is no greater than the terms $(i + \sum_{j = 0}^{i-1} \log^{(j)}\Delta + c \cdot (\Delta +1) \cdot \log^{(i)} \Delta)$ of $\tau(i, \Delta)$.\\
\textbf{Induction step:} Let $i$, $i > 1$, be an integer such that $i \leq \log^* \Delta$. Observe that $\log^{(i - 1)}\Delta > 2$.
Consider a subset $V_j$, for an index $j$, $1 \leq j \leq k^2$. Recall that the maximum degree $\Delta_j$ of the induced subgraph $G(V_j)$ is at most $\frac{\Delta}{\log^{(i-1)}\Delta}$. Also, observe that $i - 1 \leq \log^* (\log \Delta) \leq \log^* \left(\frac{\Delta}{\log^{(i-1)} \Delta}\right)$. Hence, by the induction hypothesis, the running time of the invocation Delta-Color($G(V_j)$, $\frac{\Delta}{\log^{(i-1)} \Delta}$, $i-1$, $\varphi_j$) is at most 
\begin{eqnarray*}
\tau(i-1, \frac{\Delta}{\log^{(i-1)}\Delta}) & = & i - 1 + (i - 1) \cdot c \cdot \left(\frac{\Delta}{\log^{(i-1)}\Delta}\right)^{\epsilon} + c \cdot \sum_{j = 0}^{i - 1} \frac{1}{2^j} \cdot \frac{\Delta}{\log^{(i-1)}\Delta} \\ & & +  c \cdot \sum_{j = 0}^{i-2} \log^{(j)}\left(\frac{\Delta}{\log^{(i-1)}\Delta}\right) + c \cdot \left(\frac{\Delta}{\log^{(i-1)}\Delta} + 1\right) \log^{(i-1)} \left(\frac{\Delta}{\log^{(i-1)}\Delta}\right). \\ \\
\end{eqnarray*}
Note that 
\begin{eqnarray*}
\left(\frac{\Delta}{\log^{(i-1)}\Delta} + 1\right) \cdot \log^{(i-1)}\left(\frac{\Delta}{\log^{(i-1)}\Delta}\right) & = &
\Delta - \frac{\Delta}{\log^{(i-1)}\Delta} \cdot \log^{(2i -2)}\Delta + \log^{(i-1)}\Delta - \log^{(2i-2)}\Delta \\ & & \leq \Delta + \log^{(i-1)}\Delta.
\end{eqnarray*}
Hence, 
\begin{eqnarray*}
\tau(i - 1, \frac{\Delta}{\log^{(i-1)}\Delta}) & \leq &  (i - 1) + (i - 1)\cdot c \cdot \left(\frac{\Delta}{\log^{(i-1)}\Delta}\right)^{\epsilon} + c \cdot \left( \sum_{j=0}^{i-1} \frac{1}{2^j} \cdot \frac{\Delta}{\log^{(i-1)}\Delta} + \Delta \right) \\ & & + c \cdot \left( \sum_{j=0}^{i-2} \log^{(j)}\Delta  + \log^{(i-1)}\Delta \right) .
\end{eqnarray*}
As $i \leq \log^* \Delta$, $\log^{(i-1)} \Delta > 2$. Hence
\begin{eqnarray*}
\sum_{j=0}^{i-1} \frac{1}{2^j} \cdot \frac{\Delta}{\log^{(i-1)}\Delta} +\Delta \leq \sum_{j=0}^{i-1} \frac{1}{2^j} \cdot \frac{\Delta}{2} + \Delta = \sum_{j=1}^{i} \frac{1}{2^j} \cdot \Delta + \Delta = \sum_{j=0}^{i} \frac{1}{2^j} \cdot \Delta. 
\end{eqnarray*}
Consequently, 
\begin{eqnarray*}
\tau(i - 1, \frac{\Delta}{\log^{(i-1)}\Delta}) \leq (i - 1) + (i - 1) \cdot c \cdot \Delta^{\epsilon} + c \cdot \sum_{j = 0}^{i} \frac{1}{2^j} \Delta  + c \cdot \sum_{j = 0}^{i-1} \log^{(j)} \Delta.
\end{eqnarray*}
 By Theorem \ref{reltime} ($p = \left\lfloor \log^{(i-1)} \Lambda \right \rfloor$, $q = \left\lfloor \Lambda^{\epsilon} \right \rfloor$), since the $(c \cdot \Delta^2)$-coloring $\lambda$ is computed in the very beginning of the computation, the running time $\tau_{RC}$ of Procedure Defective-Color is at most $c \cdot \Delta^{\epsilon}$. Computing an $O(d^2)$-coloring of $G(V_j)$ requires exactly one round. Recall that $d = \left\lfloor \Lambda / k \right \rfloor$, $k = \left\lfloor \log^{(i-1)} \Lambda \right \rfloor$, and $\Lambda = \Delta$. The running time $\tau_{KW}$ of the KW iterative procedure on a $(k^2 \cdot (d + 1))$-colored graph, is 
\begin{equation} \label{eq:tau}
 \tau_{KW} \leq  (\Delta + 1) \cdot \left \lceil \log \left( \frac{(d + 1) \cdot k^2}{\Delta + 1} \right) \right \rceil \leq  (\Delta + 1) \cdot \log \left \lceil \left( \frac{(\left\lfloor \Delta /k \right\rfloor + 1) \cdot \left\lfloor \log^{(i-1)} \Delta \right \rfloor^2} {\Delta + 1} \right) \right \rceil. 
\end{equation}
 
\noindent Also,
\begin{equation} \label{eq:log}
\left \lceil \log \left( \frac{ (\frac{\Delta}{ \log^{(i-1)} \Delta} + 1) \cdot \left\lfloor \log^{(i-1)} \Delta \right \rfloor^2}{\Delta + 1} \right) \right \rceil \leq \left\lceil \log \left(\log^{(i-1)} \Delta + \frac{(\log^{(i-1)} \Delta)^2}{\Delta + 1}\right) \right \rceil.
\end{equation}
For $i \geq 2$, $\frac{(\log^{(i-1)}\Delta)^2}{\Delta + 1} \leq 1$.
Hence the right-hand-side of (\ref{eq:log}) is at most $\left\lceil \log ( \log^{(i-1)} \Delta + 1) \right \rceil$.

Since $\left\lceil \log(x + 1) \right \rceil \leq 2 \cdot \log x$ for all $x \geq 2$, and also for $i \leq \log^* \Delta$, $\log^{(i-1)} \Delta \geq 2$, it follows that the right-hand-side of (\ref{eq:log}) is at most $2 \cdot \log^{(i)} \Delta$.
Hence, by (\ref{eq:tau}), $$\tau_{KW} \leq 2 \cdot (\Delta + 1) \cdot \log^{(i)} \Delta.$$ 
Hence the running time $\tau(i, \Delta)$ of Procedure Delta-Color($G$, $\Delta$, $i$, $\lambda$) satisfies $$\tau(i, \Delta) = \tau_{RC} + \tau_{KW} + \tau(i - 1, \frac{\Delta}{\log^{(i-1)}\Delta}) + 1 \leq  i + i \cdot c \cdot \Delta^{\epsilon} + c \cdot \sum_{j = 0}^{i} \frac{1}{2^j} \Delta  + c \cdot \sum_{j = 0}^{i-1}\log^{(j)} \Delta + c \cdot (\Delta + 1) \log^{(i)}\Delta.$$
}
Our final algorithm starts with invoking the SV algorithm to produce a $(c \cdot \Delta^2)$-coloring $\Lambda$ of the input graph $G$. Then it invokes procedure Delta-Color with
$\Lambda =\Delta$,
$i = \log^* \Delta$, and $\varphi = \lambda$. Theorem \ref{dcor} and Lemma \ref{druntime} imply our main result which is summarized in the following theorem.

\begin{thm} \label{deltat}
Procedure Delta-Color invoked on an input graph $G$ with a $(c \cdot \Delta^2)$-coloring  $\lambda$ computed by the SV algorithm, and with the parameter $i = \log^* \Delta$, computes a $(\Delta + 1)$-coloring in time $O(\Delta) + \frac{1}{2} \log^* n$.
\end{thm}


It is well-known \cite{L92} that given a $(\Delta + 1)$-coloring one can produce an MIS within $\Delta + 1$ rounds. Consequently, Theorem \ref{deltat} implies that our algorithm in conjunction with the reduction from \cite{L92} produces an MIS in time $O(\Delta) + \frac{1}{2} \log^* n$.

Next, we provide a tradeoff between the running time and the number of colors, and show that for any fixed value of a parameter $t$, $1 < t \leq \Delta^{1/4}$, one can achieve an $O(\Delta \cdot t)$-coloring in $O(\Delta / t) + \frac{1}{2} \log^* n$ time. This tradeoff may be useful when one needs a coloring that employs less than $\Delta^2$ colors, but cannot afford spending as much as $O(\Delta)$ time.

Set $p = t$, $q = \Delta^{3/4}$. By Theorems \ref{relcor} and \ref{reltime}, Procedure Defective-Color invoked with these parameters computes an $O(\Delta/t)$-defective $(t^2)$-coloring of $G$ in time $O(\Delta^{3/4}) + \frac{1}{2} \log^* n$. Let $\varphi$ denote the resulting coloring, and $V_j$ denote the set of vertices that were assigned the color $j$, for $1 \leq j \leq t^2$. Recall that $\Delta_j = \Delta(G(V_j)) = O(\Delta / t)$. Next, for $1 \leq j \leq t^2$, in parallel color each $G(V_j)$ with $O(\Delta/ t)$ colors using Procedure Delta-Color (Algorithm 2) using $t^2$ distinct palettes, in time $O(\Delta / t)$. (The additive term $\frac{1}{2} \log^* n$ is eliminated from the running time by employing the $O(\Delta^2)$-coloring that was computed earlier by Procedure Defective-Color, instead of recomputing it from scratch.) The resulting coloring is a legal $O(\Delta \cdot t)$-coloring, and the total running time is $O(\Delta / t + \Delta^{3/4}) + \frac{1}{2} \log^* n = O(\Delta / t) + \frac{1}{2} \log^* n$.
This result is summarized in the following theorem.
\begin{thm} \label{trade}
For a parameter $t$, $1 < t \leq \Delta^{1/4}$, our algorithm computes an $O(\Delta \cdot t)$-coloring in time $O(\Delta / t) + \frac{1}{2} \log^* n$.
\end{thm}

Next, we extend the result to hold for the entire range $1 < t \leq \Delta^{1-\epsilon}$, for an arbitrarily small constant $\epsilon$, $0 < \epsilon < 1$. In order to obtain this extended result, we eliminate the additive term of $\Delta^{3/4}$ from the running time by invoking Procedure Defective-Color several times, with parameters $p$ and $q$ that are considerably smaller than $t$ and $\Delta^{3/4}$, respectively. The extended procedure is called {\em Procedure Tradeoff-Delta-Color}. If $t < \Delta^{1/4}$, Procedure Tradeoff-Delta-Color acts as described above. Otherwise, set $p = (\min \left\{ t, \Delta / t \right \})^{1/3}$, $q = p^3 = \min \left \{t , \Delta/ t \right \}$. In the first iteration, invoke Procedure Defective-Color with the parameters $p$ and $q$ on the input graph $G$. This invocation produces a defective coloring that partitions the vertex set $V$ of $G$ into $p^2$ subsets $V_j$, such that $\Delta(G(V_j)) = O(\Delta/p)$, for $1 \leq j \leq p^2$. ($V_j$ denotes the set of vertices that were assigned the color $j$). In the second iteraton, invoke Procedure Defective-Color again with the same parameters $p$ and $q$ on all the subgraphs $G(V_j)$, for $1 \leq j \leq p^2$, in parallel. Consequently, each graph $G(V_j)$ is $O(\Delta/p^2)$-defective $(p^2)$-colored. For $1 \leq i,j \leq p^2$, let $U_i^j$ denote the set of vertices that were assigned the color $i$ by the invocation of Procedure Defective-Color on $G(V_j)$. Once the second iteration is finished, the vertex set $V$ of the input graph $G$ is partitioned into $p^4$ disjoint subsets, each inducing a subgraph with maximal degree at most $O(\Delta/p^2)$. Assigning a distinct color to each subset (e.g., the subset $U_i^j$ is assigned the color $(i-1)p^2 + j$) yields an $O(\Delta/p^2)$-defective $(p^4)$-coloring of $G$. Procedure Tradeoff-Delta-Color proceeds in this manner for $\left\lfloor \log_p t \right \rfloor$ iterations. (Observe that $\left\lfloor \log_p t \right \rfloor = O(1)$, since $p \geq \min \left \{\Delta^{1/12}, \Delta^{\epsilon/3} \right\}$, and $t < \Delta$.)

In the beginning of iteration $\ell$, $1\ \leq \ell \leq \left\lfloor \log_p t \right \rfloor$, the input graph is $O(\Delta/p^{\ell-1})$-defective $(p^{2\ell - 2})$-colored. Then Procedure Defective-Color is invoked with the parameters $p$ and $q$ on all $p^{2\ell - 2}$ subgraphs induced by the color classes, in parallel. Consequently, in the end of the iteration, the input graph $G$ is
$O(\Delta/p^{\ell})$-defective $(p^{2\ell})$-colored. Once $\left\lfloor \log_p t \right \rfloor$ iterations have been executed, the input graph is $O(\Delta/p^{\left\lfloor \log_p t \right \rfloor}$-defective $(p^{2 \left\lfloor \log_p t \right \rfloor})$-colored. An additional iteration of executing Procedure Defective-Color with the parameters $p' = t / p^{\left\lfloor \log_p t \right \rfloor}$ and $q' = q = \min \left\{t, \Delta/ t \right\}$, in parallel on all subgraphs induced by the color classes formed by the previous iteration, produces an $O(\Delta \cdot t)$-defective $(t^2)$-coloring. 
Let $Z_j$ denote the set of vertices colored $j$ by this coloring, for  $j$, $1 \leq j \leq t^2$. We apply Procedure Delta-Color on each subgraph $G(Z_j)$ in parallel. Within $O(\Delta /t) +\frac{1}{2} \log^* n$ time each $G(Z_j)$ is $O(\Delta/t)$-colored, and thus the entire graph $G$ is $O(\Delta/t) \cdot t^2$ = $O(\Delta \cdot t)$-colored.

The running time of Procedure Tradeoff-Delta-Color is analyzed in the next lemma.
\begin{lem}
For an arbitrary small constant $\epsilon$, $0 < \epsilon <1$, and an arbitrary parameter $t$, $\Delta^{1/4} \leq t \leq \Delta^{1-\epsilon}$, the running time of Procedure Tradeoff-Delta-Color is $O(\Delta/ t +  \log^* n)$.
\end{lem}
\begin{proof}
By Theorem \ref{reltime}, the running time of each iteration of Procedure Tradeoff-Delta-Color is $O(q + \log^* n) = O(\min \left\{t, \Delta/t \right\} + \log^* n)$. The number of iterations is $\left\lfloor \log_p t \right \rfloor + 1 = O(1)$, since $p = (\min\left\{t, \Delta/t \right \})^{1/3} \geq \min\left\{\Delta^{1/12}, \Delta^{\epsilon / 3} \right \}$. Hence the running time of Procedure Tradeoff-Delta-Color is $O(\min\left\{ t, \Delta/ t \right \} + \Delta/ t  + \log^* n) = O(\Delta / t + \log^* n)$.
\end{proof}

The resulting generalization of the tradeoff from Theorem \ref{trade} is summarized in the following corollary.
\begin{col} 
For an arbitrarily constant $\epsilon$, $0 < \epsilon < 1$, and an arbitrary parameter $t$, $1 < t \leq \Delta^{1-\epsilon}$, Procedure Tradeoff-Delta-Color computes an $O(\Delta \cdot t)$-coloring in time $O(\Delta / t + \log^* n)$.
\end{col}

Finally, we remark that Theorem \ref{deltat} implies an improved tradeoff for coloring graphs of bounded arboricity, and an improved algorithm for computing an MIS for the latter family of graphs. Specifically, in \cite{BE08} we have shown (Theorem 5.1) that graphs of arboricity at most $a$ can be $O(a \cdot t)$-colored in time $O(\frac {a}{t} \log n + a \log a)$, for any parameter $t$, $1 \leq t \leq a$. The algorithm that achieves this tradeoff employs the KW algorithm on graphs wiht maximum degree $O(a)$. This step requires $O(a \log a + \log^* n)$ time. By replacing the invocation of the KW algorithm by an invocation of our new algorithm from Theorem \ref{deltat} we improve the running time of this step to $O(a + \log^* n)$, and the overall running time to $O(\frac {a}{t} \log n + a)$.

In addition in \cite{BE08} we used this tradeoff to achieve an algorithm for computing MIS for graphs with arboricity at most $a$ in time $O(a \sqrt {\log n} + a \log a)$. (See Theorem 6.4 in \cite{BE08}.) This is done by first computing the $O(a \cdot t)$-coloring, and then converting the $O(a \cdot t)$-coloring into MIS within additional $O(a \cdot t)$ rounds. By employing our improved tradeoff for $O(a \cdot t)$-coloring (in time $O(\frac{a}{t} \log n + a))$, we obtain overall time of $O(\frac{a}{t} \log n + a \cdot t)$. Finally, we set $t = \sqrt{\log n}$ and obtain the running time of $O(a \sqrt{\log n})$.

\begin{col}
For a parameter $t$, $1 \leq t \leq a$, an algorithm from \cite{BE08} that uses Procedure Delta-Color instead of the KW algorithm as a subroutine computes an $O(a \cdot t)$-coloring for graphs with arboricity at most $a$. Its running time is $O(\frac{a}{t} \log n + a)$. As a result, we can compute an MIS for this family of graphs in time $O(a \sqrt{\log n})$.
\end{col}


\clearpage
\pagenumbering{roman}
\appendix
\centerline{\LARGE\bf Appendix}
\section{Figures}
\begin{fig}
The depicted graph is $2$-defective $2$-colored. The vertex $v$ is colored black, and it has two black neighbors. The vertex $z$ is colored white, and it has two white neighbors. All other vertices have one or less neighbors colored by the same color.
\end{fig}
\includegraphics{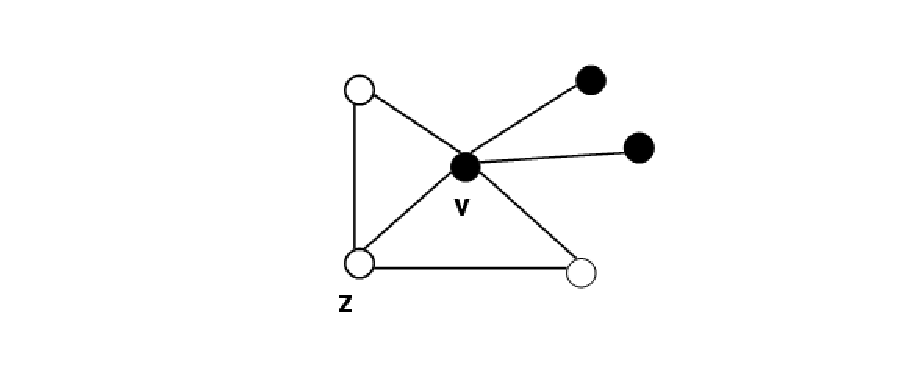}
\clearpage
\begin{fig}
An execution of Procedure Defective-Color with the parameters $p$ and $q$, such that $q = 3p^2$, on an initially $(6 q)$-colored graph $G$. Each oval represents a subgraph. The range inside the oval represents the color palette employed by the subgraph. For $j$, $1 \leq j \leq 6$, the set $V_j$ changes after each iteration, and contains all vertices that are currently colored using the palette $\left\{ (j-1) \cdot q + 1, (j-1) \cdot q + 2, ..., j \cdot q \right \}$.
\end{fig}
\includegraphics{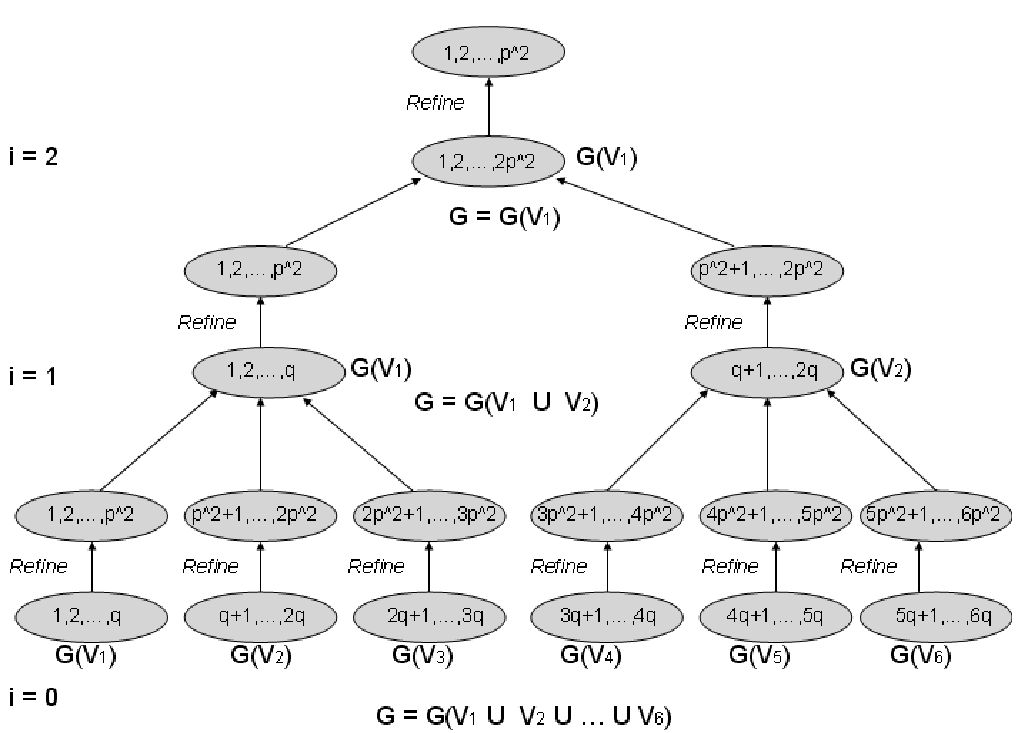}
\clearpage
\section{Some Proofs}
\APPb 
\hfill \ensuremath{\Box} \\ \\
\noindent \textbf{Proof of Lemma 3.4:}
\APPc 
\hfill \ensuremath{\Box} \\ \\
\noindent \textbf{Proof of Lemma 4.5:}
\APPd 
\hfill \ensuremath{\Box} \\

\end{document}